\documentclass[
aps, 
prl,   
reprint,
groupedaddress,
superscriptaddress,
longbibliography
]{revtex4-2}

\usepackage{indentfirst}
\usepackage{graphicx}
\usepackage{dutchcal}
\usepackage{bm}
\usepackage{amsmath,amssymb}
\usepackage{physics}
\usepackage{xcolor}
\usepackage[
colorlinks=true,
linkcolor=blue!45!black,
citecolor=blue!45!black,
urlcolor=blue!45!black,
filecolor=blue!45!black,
pdfborder={0 0 0}
]{hyperref}

\begin{document}
	
	\title{Acoustic quantum skyrmion--valley Hall effect}
	
	\author{Lei Liu}
	\affiliation{National Laboratory of Solid State Microstructures and Department of Materials Science and Engineering, Nanjing University, Nanjing 210093, China}
	
	\author{Xiujuan Zhang\thanks{Corresponding author}}
	\email[]{xiujuanzhang@nju.edu.cn}
	\affiliation{National Laboratory of Solid State Microstructures and Department of Materials Science and Engineering, Nanjing University, Nanjing 210093, China}
	
	\author{Ming-Hui Lu}
	\affiliation{National Laboratory of Solid State Microstructures and Department of Materials Science and Engineering, Nanjing University, Nanjing 210093, China}
	\affiliation{Jiangsu Key Laboratory of Artificial Functional Materials, Nanjing 210093, China}
	\affiliation{Collaborative Innovation Center of Advanced Microstructures, Nanjing University, Nanjing 210093, China}
	
	\author{Yan-Feng Chen}
	\affiliation{National Laboratory of Solid State Microstructures and Department of Materials Science and Engineering, Nanjing University, Nanjing 210093, China}
	\affiliation{Collaborative Innovation Center of Advanced Microstructures, Nanjing University, Nanjing 210093, China}
	
	\begin{abstract}
		Skyrmions are particle-like topological textures that hold great promise for low-power electronics and wave-based functionalities. Yet their utility is hindered by the lack of robust and controllable transport. Here, we show that band topology can be harnessed to overcome this limitation. We experimentally realize an acoustic quantum skyrmion--valley Hall effect in a surface phononic crystal via engineered spin--orbit--momentum interaction. Skyrmions emerge as valley-locked topological edge states, robustly propagating along designed domain walls. Crucially, the skyrmion transport exhibits concurrent orbital angular momentum (OAM)--valley locking and spin--texture locking, enabling controllable propagation through selective excitation. Our results establish a direct correspondence between real-space and momentum-space topology, providing a general strategy for robust, controllable skyrmion transport.
	\end{abstract}
	
	
	\maketitle
	\textit{Introduction}---Magnetic skyrmions are topologically protected swirling spin textures, whose nanoscale size and topological stability endow them with great potential for low-power, high-density storage and computing technologies \cite{bogdanov1989thermodynamically,muhlbauer2009skyrmion,romming2013writing,jiang2015blowing,fert2017magnetic,bogdanov2020physical,reichhardt2022statics}. Recently, the concept of skyrmions has been extended to classical wave platforms, including optics \cite{science.aau0227,du2019deep,science.aba6415}, acoustics \cite{PhysRevLett.127.144502,sciadv.adf3652,liu2025acoustic}, and water waves \cite{wang2025topological}, where they emerge in various vector fields such as electric polarization, elastic displacement and fluid velocity. These developments broaden the physical scope of skyrmions and open practical avenues for skyrmion-enabled wave functionalities, such as topological lasing \cite{PhysRevResearch.3.023055,Kerridge-Johns:24}, computing \cite{wang2025perturbation} and sensing \cite{chen2025programmable}.
	
	Despite these advances, a universal challenge remains across different systems in practical applications, namely the lack of robust and controllable skyrmion transport. In magnetic systems, the dynamics of skyrmions in tracks are perturbed by the skyrmion Hall effect, which causes a transverse motion relative to the driving force \cite{jiang2017direct,litzius2017skyrmion}. This poses a major challenge for devices \cite{gobel2021beyond}, as skyrmions can drift toward the track edges and annihilate. In wave systems, many realizations rely on static interference patterns of surface waves which inherently lack a mechanism for long-range propagation \cite{science.aau0227,du2019deep,science.aba6415,PhysRevLett.127.144502,wang2025topological,sciadv.adf3652,chen2025programmable}. Recently, skyrmionic beams have been demonstrated in free space \cite{PhysRevA.102.053513,shen2021supertoroidal,teng2025construction,fan2025topological,Zhen26}, yet their topological textures inevitably deform during paraxial propagation due to the Gouy phase effect \cite{PhysRevA.102.053513,teng2025construction,fan2025topological}. Although these structural distortions can be externally compensated, the required complex optical architectures are bulky and lack integrability \cite{Zhen26}. Skyrmion transport has also been observed in artificially designed Archimedes metastructures, but the reliance on trivial geometric couplings renders the textures susceptible to scattering and distortion \cite{sun2024acoustic}.
	
	\begin{figure}[!b]
		\centering
		\includegraphics[width=1\linewidth]{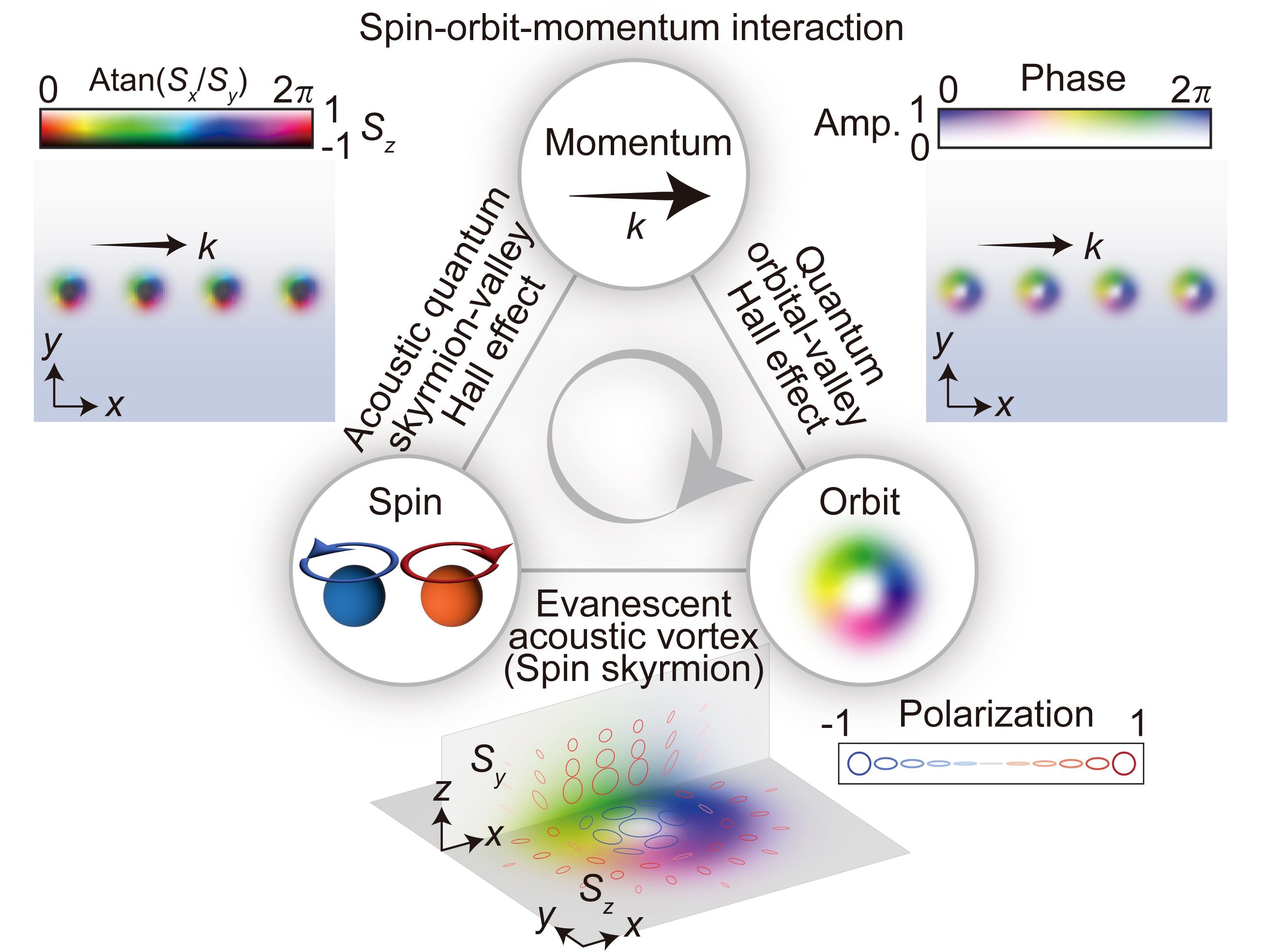}
		\caption{\label{fig1}
			Mechanism of the acoustic quantum SVH effect enabled by spin--orbit--momentum interaction.
		}
	\end{figure}
	Here, we experimentally demonstrate robust and controllable skyrmion transport by harnessing the synergy between momentum-space band topology and real-space topological textures through a quantum skyrmion--valley Hall (SVH) effect. This is enabled by the interplay of lattice symmetry and spin--orbit interaction in an engineered surface phononic $p$-orbital honeycomb lattice. By superposing $p_x$ and $p_y$ orbitals at each lattice site, we synthesize an internal orbital angular momentum (OAM) degree of freedom coupled to the lattice momentum, thereby establishing a quantum orbital--valley Hall (OVH) topology. It supports valley-polarized topological edge states carrying nonzero vortex charges (Fig.~\ref{fig1}). By embedding this architecture in an evanescent acoustic field, the spin--orbit--momentum interaction effectively maps these vortex edge states into spin skyrmions. Within this framework, the skyrmions are simultaneously manifested as real-space topological spin textures and are locked to momentum-space valley topology. As a result, their propagation exhibits both OAM--valley locking and spin--texture locking, providing a dual and programmable control mechanism. By establishing a direct correspondence between real-space and momentum-space topology, our approach offers a general strategy to mitigate skyrmion drift and achieve robust, controllable transport, with immediate relevance to a broad class of skyrmionic wave and condensed-matter platforms.
	
\begin{figure}[!b]
		\centering
		\includegraphics[width=1\linewidth]{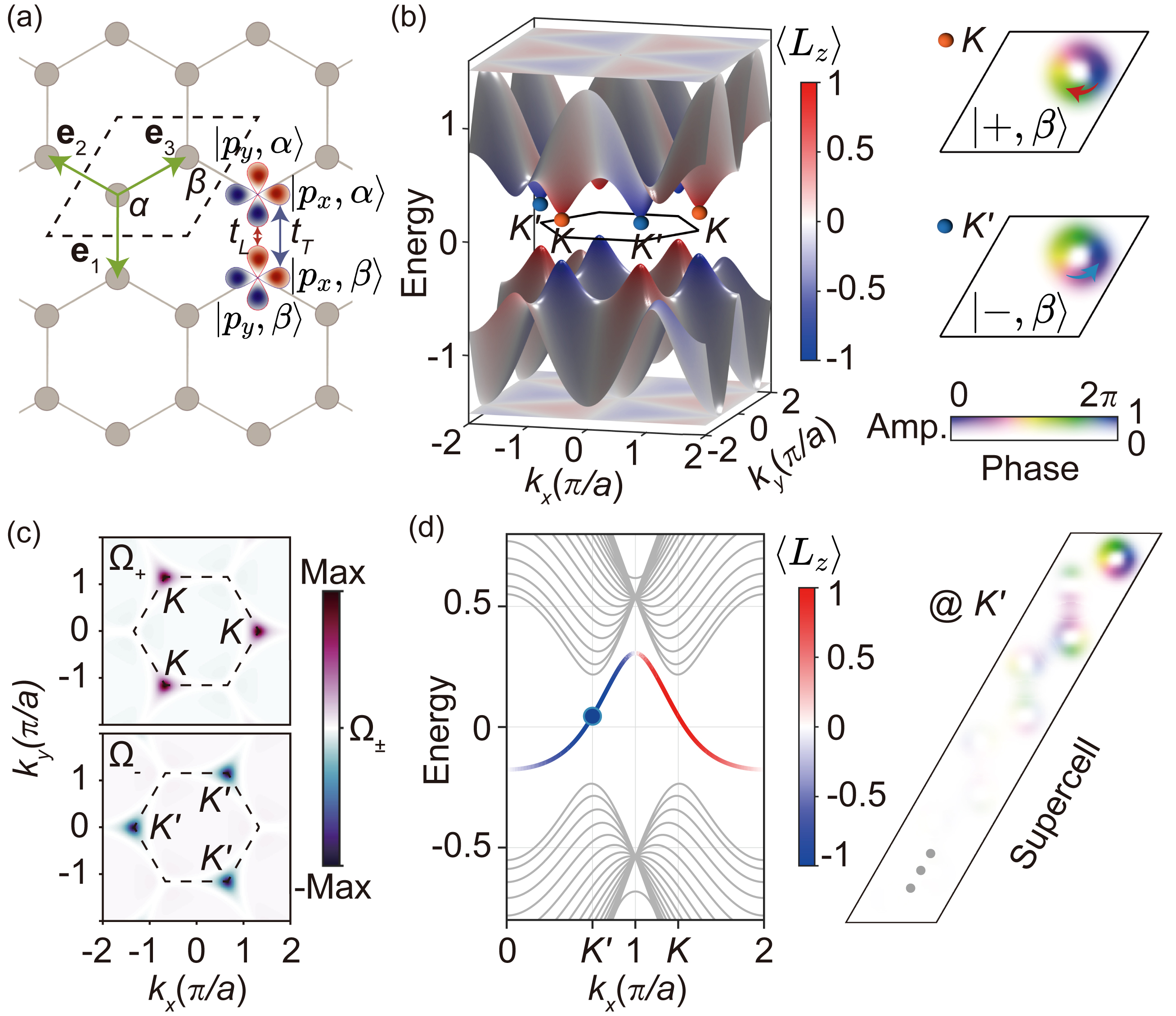}
		\caption{\label{fig2}
			(a) Schematic of the $p$-orbital honeycomb lattice.
			(b) Band structure colored by the OAM polarization $\langle L_z\rangle$, with the eigenstates at the $K$ and $K'$ valleys shown on the right.
			(c) OAM-projected Berry curvatures.
			(d) Left: Band structure of a zigzag ribbon, with the edge dispersion colored by the edge OAM polarization. Right: Edge-state wavefunction at the $K'$ valley.
		}
	\end{figure}

	\textit{Orbital--valley Hall topology}---We begin with a honeycomb lattice in which each sublattice site
	$\mu\in\{\alpha,\beta\}$ hosts two degenerate in-plane $p$ orbitals $\ket{p_x,\mu}$ and $\ket{p_y,\mu}$
	[Fig.~\ref{fig2}(a)]. The system is governed by transverse ($t_T$) and longitudinal ($t_L$) couplings with $t_T\ll t_L$, and we set $t_T=0$ without loss of generality. It is convenient to switch the orbital basis into the OAM basis via $\ket{\pm,\mu}=\frac{1}{\sqrt2}\big(\ket{p_x,\mu}\pm i\ket{p_y,\mu}\big)$, where $\ket{\pm,\mu}$ are OAM eigenstates with quantum numbers $l_z=\pm1$ on sublattice $\mu$. We introduce Pauli matrices $\tau_i$ and $\sigma_i$ ($i=x,y,z$) for the OAM and sublattice degrees of freedom, respectively, and denote the corresponding identity matrices by $\tau_0$ and $\sigma_0$. Then, the Bloch Hamiltonian can be written as
	$H(\mathbf{k})=\frac{t_L}{2}
	\begin{pmatrix}
		H_{+}(\mathbf{k}) & H_{\mathrm{OOC}}(\mathbf{k})\\
		H_{\mathrm{OOC}}^{\dagger}(\mathbf{k}) & H_{-}(\mathbf{k})
	\end{pmatrix}$ (Supplementary Sec.~1 \cite{supp}), 
	in the basis $(\ket{+,\alpha},\ket{+,\beta},\ket{-,\alpha},\ket{-,\beta})$.
	$H_{\pm}(\mathbf k)$ describe graphene-type hopping within each OAM subspace and take the form
    $H_{+}(\mathbf k)=H_{-}(\mathbf k)=f(\mathbf k)\,\sigma_{+}+f^{*}(\mathbf k)\,\sigma_{-},$
	where $f(\mathbf k)=\sum_{n=1}^{3}e^{i\mathbf k\cdot \mathbf e_n}$, $\mathbf e_n$ are the nearest-neighbor bond vectors, and $\sigma_{\pm}=(\sigma_x\pm i\sigma_y)/2$ are sublattice ladder operators.
	$H_{\mathrm{OOC}}(\mathbf k)$ represents a bond-direction-dependent orbit--orbit coupling (OOC) and is given by
	$H_{\mathrm{OOC}}(\mathbf k)=g(\mathbf k)\,\sigma_{+}+g(-\mathbf k)\,\sigma_{-},$
	with $g(\mathbf k)=\sum_{n=1}^{3}e^{i\mathbf k\cdot \mathbf e_n}e^{i2\theta_n}$, where $\theta_n$ is the azimuthal angle of the $n$th bond.
	$H_{\mathrm{OOC}}(\mathbf k)$ couples the OAM degree of freedom to the lattice momentum, giving rise to an orbit--momentum interaction. Enforced by the lattice $C_3$ symmetry, this interaction is valley-dependent, yielding a nontrivial valley topology (Supplementary Sec.~2 \cite{supp}).
	
	To enable valley-locked transport, we break sublattice (inversion) symmetry by introducing a staggered sublattice potential $H_m=\Delta m\,\tau_0\otimes\sigma_z$, which assigns opposite onsite potentials $+\Delta m$ and $-\Delta m$ to the $\alpha$ and $\beta$ sublattices, respectively.
	This symmetry breaking induces a nonzero OAM polarization, $\langle L_z\rangle=\langle\psi|L_z|\psi\rangle$, where $L_z=\tau_z\otimes\sigma_0$ is the OAM operator and $\ket{\psi}$ is the Bloch eigenstate (Supplementary Sec.~3 \cite{supp}).
	Taking $t_L=1$ and $\Delta m=-0.2$, Fig.~\ref{fig2}(b) shows the resulting band structure incorporating $\langle L_z\rangle$.
	The staggered potential opens a gap at the valleys, yielding fully OAM-polarized eigenstates with $\langle L_z\rangle=\pm1$, thereby establishing OAM--valley locking.
	In the third band (i.e., the higher-energy band), the eigenstates are predominantly localized on the $\beta$ sublattice, with the state $\ket{+,\beta}$ locked to the $K$ valley and $\ket{-,\beta}$ to $K'$ under time-reversal symmetry. In contrast, the second (lower-energy) band hosts eigenstates localized on the $\alpha$ sublattice, exhibiting opposite OAM polarization (Supplementary Sec.~4 \cite{supp}).      
	
	The topology of this bandgap is characterized by the OAM-projected Berry curvature. In analogy with the spin-projected Berry curvature \cite{PhysRevB.80.125327,PhysRevLett.107.066602,deng2020acoustic}, we resolve the gap-isolated band subspace into positive ($+$) and negative ($-$) OAM sectors through the projected OAM operator, and evaluate the corresponding Berry curvatures $\Omega_{\pm}$ (Supplementary Sec.~5 \cite{supp}).
	As shown in Fig.~\ref{fig2}(c), $\Omega_{+}$ and $\Omega_{-}$ are peaked at $K$ and $K'$, respectively, confirming the OVH topological phase. By bulk--edge correspondence, this momentum-space topology implies OAM--valley-locked edge states.
	This is corroborated by the band structure of a zigzag ribbon, which exhibits in-gap edge states, as shown in the left panel of Fig.~\ref{fig2}(d), in accordance with the edge OAM polarization $\langle L_z\rangle$ (calculated locally at the boundary site; see Supplementary Sec.~6 \cite{supp}) and the edge-state wavefunction (see the right panel for the example at the $K'$ valley, revealing an edge-localized vortex with $l_z=-1$). 
	
	\textit{Acoustic quantum skyrmion--valley Hall effect}---In acoustics, the OAM arises from the twisted azimuthal phase gradient of the scalar pressure field $P$, whereas the spin angular momentum (SAM) $\mathbf s$ describes the local rotation of the velocity vector field $\mathbf v$ and is defined as $\mathbf s=\frac{\rho}{2\omega}\Im\!\left(\mathbf v^{*}\times\mathbf v\right)$, where $\rho$ is the mass density and $\omega$ the angular frequency. Through the linearized Euler equation, the vector velocity field is intrinsically linked to the gradient of the scalar pressure field, thereby promoting spin--orbit interaction in structured acoustic fields \cite{advs.202409377}.
	
	It has been previously demonstrated that engineering spin--orbit interaction in an evanescent field enables the conversion of acoustic vortices with $l_z=\pm1$ into N\'eel-type spin skyrmions with skyrmion number $n_{\rm sk}=\pm1$ \cite{liu2025acoustic}. 
	Building on this principle, we incorporate the above tight-binding vortex lattice into a surface phononic crystal that supports out-of-plane evanescent fields, thereby enabling the formation and transport of skyrmionic textures inheriting the OVH topology.
	
\begin{figure}[!b]
		\centering
		\includegraphics[width=1\linewidth]{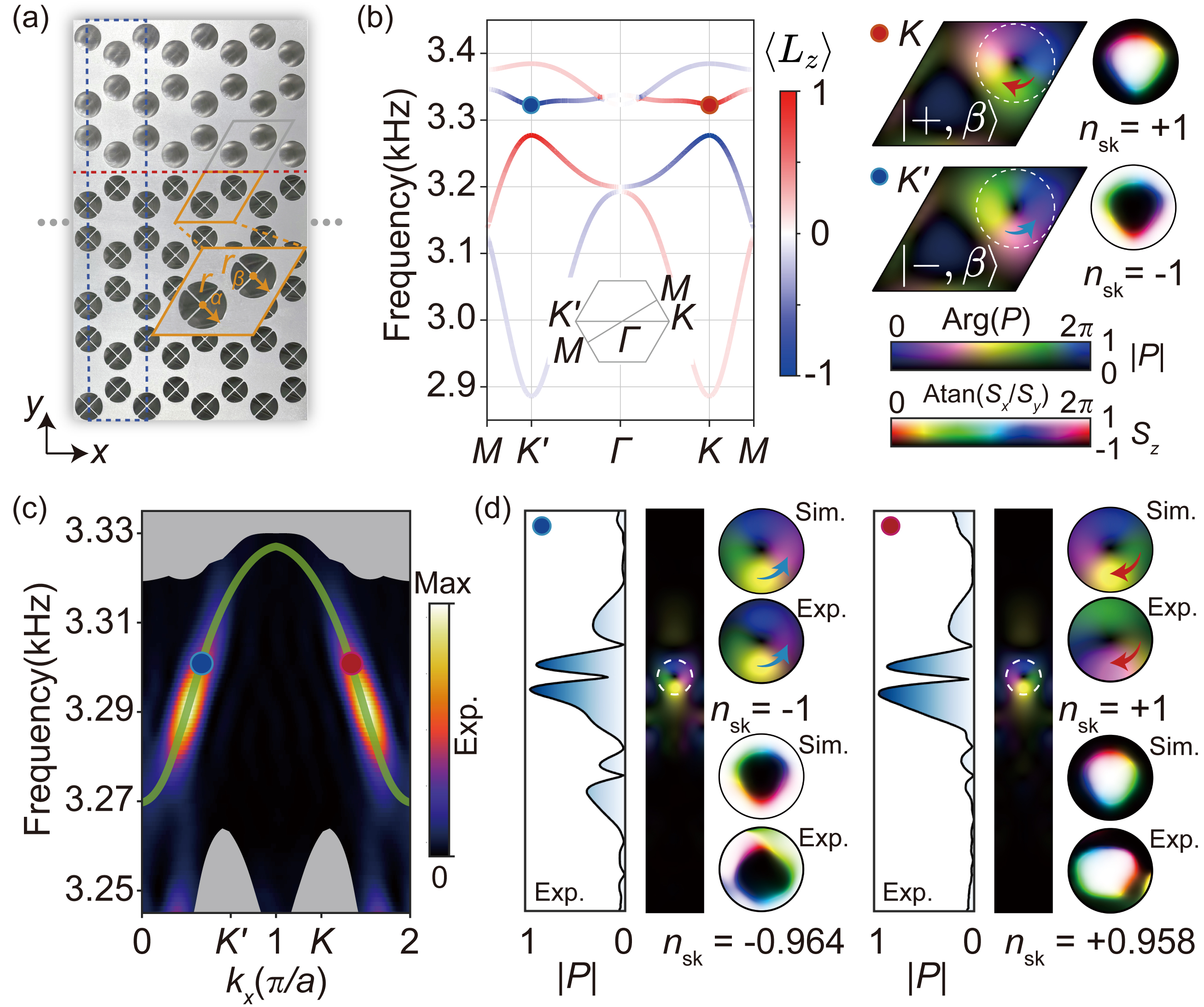}
		\caption{\label{fig3}
			(a) Fabricated surface phononic crystals. The blue dashed line outlines the supercell.
			(b) Left: Band structure colored by the normalized $\langle L_z\rangle$. Right: Pressure field distributions of the valley eigenstates and the corresponding spin textures within the $\beta$ cavity.
			(c) Measured edge dispersion, with numerically calculated bulk (gray) and edge (green) states.
			(d) Measured pressure and spin distributions of two skyrmion edge states marked in (c).
		}
	\end{figure}

	The designed phononic crystal sample is shown in Fig.~\ref{fig3}(a), comprising open resonant cavities patterned on a steel plate forming a honeycomb lattice. In the OVH domain, each sublattice site ($\alpha$ or $\beta$) consists of four petaloid cavities that support two degenerate $p$-orbital-like dipole modes \cite{liu2025acoustic}. By choosing different radii for the $\alpha$ and $\beta$ cavities, sublattice symmetry is broken, thereby inducing nonzero OAM polarization. To quantify the OAM polarization in this acoustic system, we integrate the OAM density over the two sublattice cavities as   
    $\langle L_z\rangle=\sum_{\mu=\alpha,\beta}\iint_{S_\mu}\!\big(\mathbf r_\mu\times \mathbf p\big)_z\,\mathrm dS,$
	where $\mathbf p=\frac{\rho}{2\omega}\Im\!\left[\mathbf v^{*}\!\cdot(\nabla)\mathbf v\right]$ is the canonical momentum density, $S_\mu$ denotes the $\mu$ cavity surface area, and $\mathbf r_\mu$ is the position vector measured from the corresponding cavity center.
	Figure~\ref{fig3}(b) shows the band structure color-coded by the normalized $\langle L_z\rangle$. Fully OAM-polarized eigenstates with $\langle L_z\rangle=\pm1$ emerge at the valleys. In the third band, the eigenstates are localized on the $\beta$ sublattice, with $\ket{+,\beta}$ and $\ket{-,\beta}$ locked to the $K$ and $K'$ valleys, respectively. This is consistent with the tight-binding result in Fig.~\ref{fig2}(b) and indicates the OVH topological phase. Owing to the spin--orbit interaction in the evanescent acoustic wave, these vortex states are converted into spin skyrmionic textures. The real-space topology of a skyrmion is characterized by the skyrmion number
	$n_{\rm sk}=\frac{1}{4\pi}\iint \hat{\mathbf s}\cdot\left(\partial_x\hat{\mathbf s}\times\partial_y\hat{\mathbf s}\right)\,\mathrm dx\,\mathrm dy,$
	where $\hat{\mathbf s}=\mathbf s/|\mathbf s|$ is the normalized spin vector.
	The calculated $\hat{\mathbf s}$ fields (derived from the velocity field), shown in the rightmost panel of Fig.~\ref{fig3}(b), exhibit corresponding skyrmions with $n_{\rm sk}=+1$ and $-1$ for the states $\ket{+,\beta}$ and $\ket{-,\beta}$, respectively, signaling the SVH topological phase.
	
	The emergence of the SVH phase manifests an ingenious interplay between momentum-space band topology and real-space topological textures. Such a synergy, mediated by the spin--orbit--momentum interaction, endows the skyrmion edge states with a dual locking mechanism governing both OAM and SAM. To demonstrate these edge states, a topological domain wall is constructed by adjoining the SVH lattice to a trivial one [see Fig.~\ref{fig3}(a) and Supplementary Sec.~7 \cite{supp}]. Experimental evidence for these edge states is provided by the measured band structure in Fig.~\ref{fig3}(c), which exhibits an in-gap edge dispersion.
	The spatial localization of these edge states is corroborated by the measured pressure field distributions shown in Fig.~\ref{fig3}(d). Their dual characteristics are further confirmed by the nontrivial OAM polarization $\langle L_z\rangle=-1$ ($+1$) and the complementary skyrmion number of $n_{\rm sk}=-0.964$ ($+0.958$) for the mode near the $K'$ ($K$) valley.
	
	\textit{OAM--valley-locked and spin--texture-locked skyrmion transport}---Building upon the established acoustic quantum SVH effect, the skyrmion edge states exhibit a hierarchical dual-locking mechanism encompassing both OAM and SAM, which offers two distinct levels of transport controllability. At the macroscopic level, the transport is strictly OAM--valley-locked. Because the edge states at different valleys carry opposite OAM, their propagation direction can be deterministically routed by excitation sources with tailored OAM. On a finer, subwavelength scale, the transport is dictated by spin--texture locking. Owing to the intricate real-space topological spin textures of skyrmions, counterpropagating modes possess sharply distinct local spin distributions. This characteristic enables highly selective excitation and precision manipulation of skyrmion transport using sources with engineered spin configurations.
	
	\begin{figure}[!b]
		\centering
		\includegraphics[width=1\linewidth]{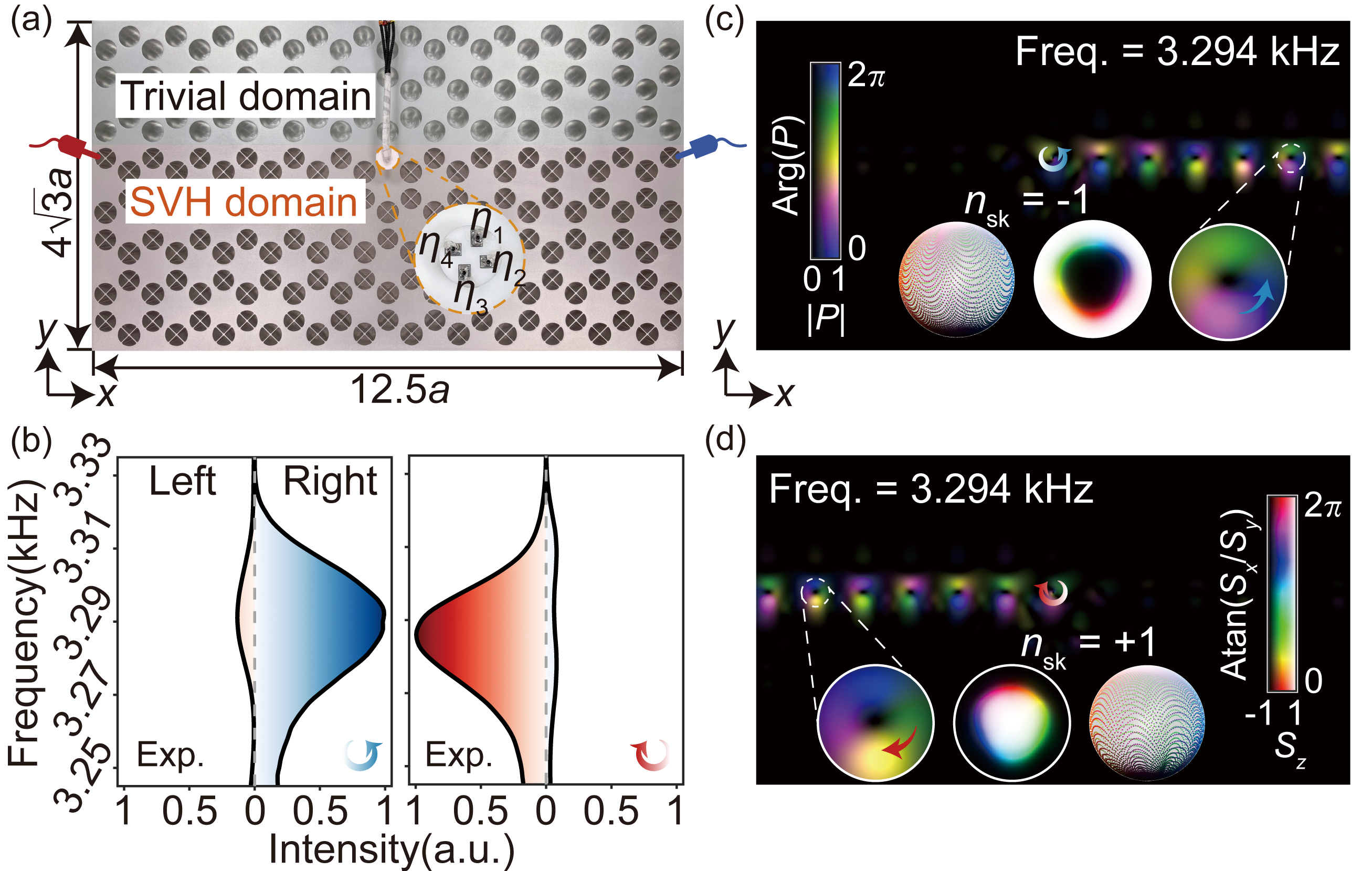}
		\caption{\label{fig4}
			(a) Experimental setup. 
			(b) Measured transmission spectra. 
			(c) Simulated pressure field distribution under OAM excitation with $l_z=-1$ at $3.294\,\mathrm{kHz}$. The zoomed-in view highlights the OAM feature with $l_z=-1$ and a skyrmion with $n_{\rm sk}=-1$ (as confirmed by the spin vector field wrapping the unit spin sphere once).
			(d) Same as (c), but for excitation with $l_z=+1$.
		}
	\end{figure}
	We first demonstrate the OAM--valley locking of skyrmion transport by showing that excitations with OAM $l_z=\pm1$ launch the skyrmion edge state in opposite directions. The OAM source is synthesized using four loudspeakers with an azimuthal phase winding and placed at the center of the domain wall [Fig.~\ref{fig4}(a)]. Figure~\ref{fig4}(b) shows the normalized transmission spectra measured at the two ends of the domain wall under excitation with $l_z=\pm1$, revealing pronounced OAM--valley-locked unidirectional transport. 
	For $l_z=-1$, the transmitted intensity is concentrated at the right end and remains negligible at the left end, indicating predominantly forward transport, whereas the propagation direction reverses for $l_z=+1$.
	To further visualize the unidirectional propagation, Figs.~\ref{fig4}(c-d) show the field distributions at an excitation frequency of $3.294\,\mathrm{kHz}$ for $l_z=\pm1$. 
	Under $l_z=-1$ excitation [Fig.~\ref{fig4}(c)], the skyrmion edge state propagates forward carrying OAM $l_z=-1$ and skyrmion number $n_{\rm sk}=-1$, as confirmed by the pressure field and spin distributions. Under $l_z=+1$ excitation [Fig.~\ref{fig4}(d)], the pressure field and spin texture reveal a backward-propagating skyrmion edge state with $l_z=+1$ and $n_{\rm sk}=+1$. These results are consistent with the measured transmission spectra. Notably, due to the valley-locking property, such edge transport can be routed along designed paths respecting $C_3$ lattice symmetry, enabling symmetry-protected and programmable skyrmion propagation (Supplementary Sec.~8 \cite{supp}).   
	
	\begin{figure}[!b]
		\centering
		\includegraphics[width=1\linewidth]{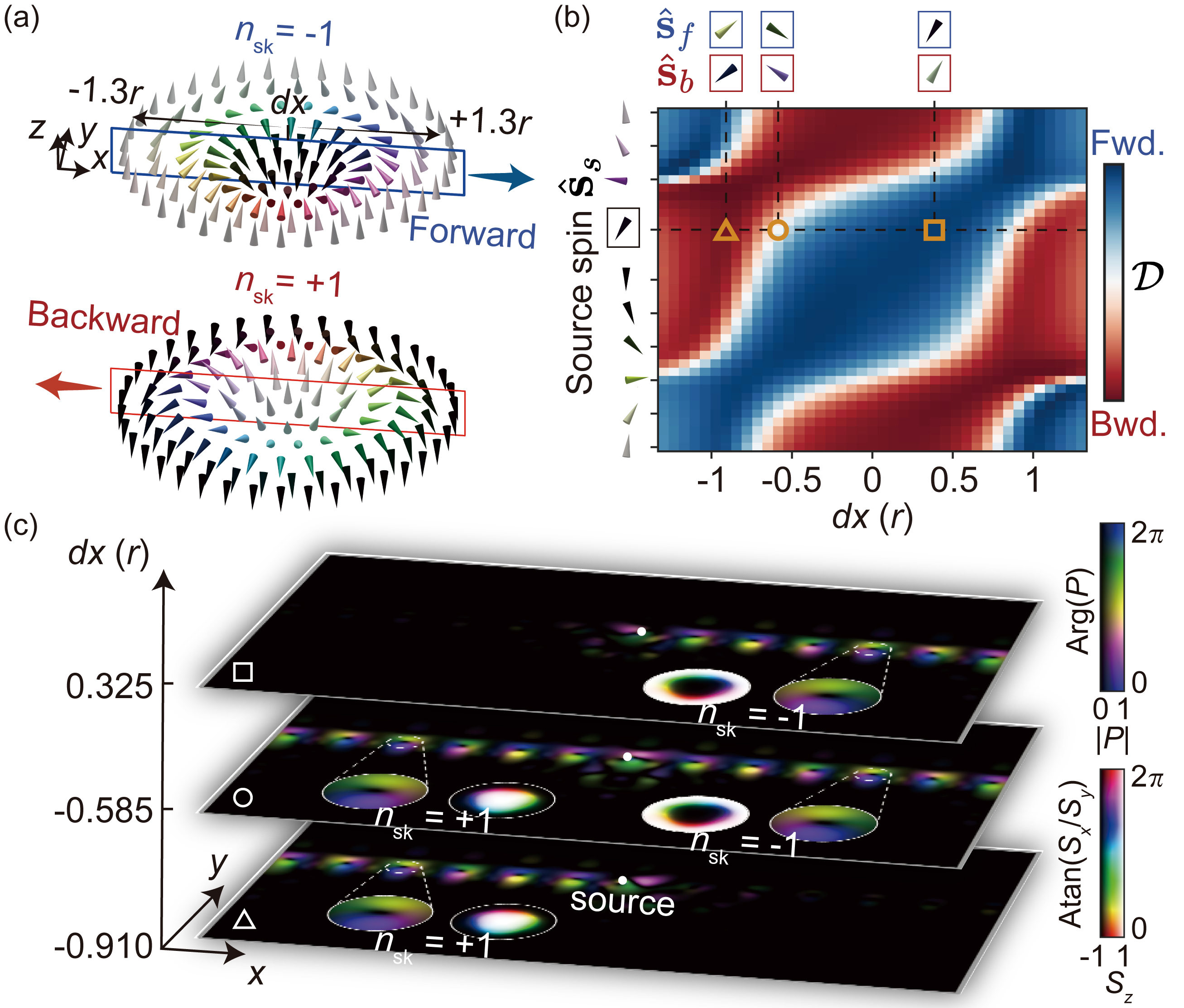}
		\caption{\label{fig5}
			(a) Spin textures carried by the two counterpropagating skyrmion edge states at $3.294\,\mathrm{kHz}$.
			(b) Simulated directional contrast $\mathcal D$ for different source spins and positions, with the local spins of the forward and backward edge states shown above.
			(c) Simulated field distributions excited by the spin source marked in (b). Zoomed-in views highlight the OAM and skyrmion features.
		}
	\end{figure}
    Beyond the OAM--valley locking, skyrmion transport further exhibits spin--texture locking on a finer scale. To elucidate this mechanism, we revisit the two counterpropagating skyrmion edge states with $n_{\rm sk}=-1$ (locked to the forward propagation) and $n_{\rm sk}=+1$ (locked to the backward propagation) at the same frequency as in Figs.~\ref{fig4}(c-d), but now excited by sources with tailored spin. Figure~\ref{fig5}(a) presents the local spin distributions of these modes, which exhibit two distinct swirling vector textures. When the source is tailored to match the local spin orientation and spatial location of a given edge state, selective excitation can be achieved, enabling precise control over not only the propagation direction but also the energy partition. This behavior is captured by Fermi's golden rule \cite{PhysRevLett.128.073602}, which gives the excitation probabilities
    of the forward and backward modes as $\Gamma_{f/b}^{(\mathrm{th})}\propto (1+\hat{\mathbf s}_{s}\cdot\hat{\mathbf s}_{f/b})/2$ (Supplementary Sec.~9 \cite{supp}), where $\hat{\mathbf s}_{s}$ denotes the source spin and $\hat{\mathbf s}_{f/b}$ the local spin of the forward/backward skyrmion edge states. In simulations, this is rigorously quantified by evaluating the transmitted powers at distinct output ports under $\hat{\mathbf s}_{s}$-excitation, yielding $\Gamma_{f/b}^{(\mathrm{sim})}\propto \iint_{S_{f/b}} \Re\!\big(P v_x^{*}\big)/2\,\mathrm dS$, where $S_{f/b}$ denote the right and left ports. With $\Gamma_{f/b}^{(\mathrm{sim})}$, the directional contrast can be further obtained via $\mathcal D=(\Gamma_f-\Gamma_b)/(\Gamma_f+\Gamma_b)$, signifying the directionality. Figure~\ref{fig5}(b) quantitatively maps the correlation between the source spin, its location, and the resulting skyrmion directionality, revealing a fine-grained dependence.
    
    A closer visualization of the position- and spin-dependent directional skyrmion transport is provided in Fig.~\ref{fig5}(c), where a source with spin $\hat{\mathbf s}_{s}=(-0.52,0.08,-0.85)$ is launched at varying positions $dx$. It is shown that as $dx$ changes, the overlap between $\hat{\mathbf s}_{s}$ and $\hat{\mathbf s}_{f/b}$ is continuously modulated, leading to distinct excitation scenarios. Specifically, at $dx=0.325r$, the source spin matches $\hat{\mathbf s}_{f}$, resulting in predominantly forward-propagating skyrmions carrying $l_z=-1$ and $n_{\rm sk}=-1$. In contrast, at $dx=-0.910r$, the same source preferentially couples to $\hat{\mathbf s}_{b}$, giving rise to reversed, backward propagation with $l_z=+1$ and $n_{\rm sk}=+1$. Interestingly, at an intermediate position ($dx=-0.585r$), the source can simultaneously couple to both modes, enabling nearly symmetric excitation and an even redistribution of energy between the two propagation directions. This principle naturally extends to arbitrary energy partition by carefully choosing the source position or engineering its spin. 
         
    It is worth noting that the demonstration of spin--texture-locked skyrmion transport is based on numerical simulations. This is primarily due to the deep-subwavelength variation of the skyrmion spin texture in real space, which lies beyond the spatial resolution of the current acoustic measurement setup. Notably, however, this pronounced deep-subwavelength spin inhomogeneity, while experimentally challenging to resolve, highlights the strong potential for high-precision displacement and spin-sensitive sensing applications.
	
	\textit{Conclusions}---We have demonstrated an acoustic quantum SVH effect by harnessing the interplay between momentum-space band topology and real-space vectorial topology. This coupling anchors skyrmions to symmetry-protected edge transport channels, enabling skyrmions to propagate robustly and deterministically along designed paths. Given the universality of lattice symmetries, our approach provides a general route to robust and controllable skyrmion manipulation, with immediate applicability to magnetic and wave-based skyrmionic systems.
	
	Built on engineered spin--orbit--momentum interaction, the SVH effect further endows skyrmion transport with OAM--valley locking and spin--texture locking, providing two complementary levels of control. In particular, the spin-resolved control operates at deep-subwavelength scales, offering a powerful mechanism for high-precision metrology, high-throughput computing, and flexible particle manipulation.
	
	More broadly, our work establishes a general framework for steering real-space topological textures via momentum-space topology. It can be naturally extended to other momentum-space topological phases (such as Chern, Floquet, and higher-order systems) and real-space vectorial degrees of freedom (including magnetization, polarization, and displacement), yielding richer topological textures and transport behaviors across diverse physical settings.
	
	\par\vspace{1\baselineskip} 
	\raggedbottom 
	\begin{acknowledgments}
		\textit{Acknowledgments}---This work is supported by the National Key R\&D Program of China (Grants No. 2023YFA1407700 and No. 2023YFA1406904).
	\end{acknowledgments}
	\par\vspace{1\baselineskip} 
	\textit{Data availability}---The data that support the findings of this study are openly available \cite{liu_2026_18328890}.
	\bibliography{references}

@article{bogdanov1989thermodynamically,
	author = {Bogdanov, A. and Yablonskiui, D.},
	year = {1989},
	month = {01},
	pages = {101},
	title = {Thermodynamically stable ``vortices'' in magnetically ordered crystals. {The} mixed state of magnets},
	volume = {68},
	journal = {Sov. Phys. JETP}
}

@article{muhlbauer2009skyrmion,
	author = {S. Mühlbauer  and B. Binz  and F. Jonietz  and C. Pfleiderer  and A. Rosch  and A. Neubauer  and R. Georgii  and P. Böni },
	title = {Skyrmion Lattice in a Chiral Magnet},
	journal = {Science},
	volume = {323},
	number = {5916},
	pages = {915-919},
	year = {2009},
	doi = {10.1126/science.1166767},
}

@article{romming2013writing,
	author = {Niklas Romming  and Christian Hanneken  and Matthias Menzel  and Jessica E. Bickel  and Boris Wolter  and Kirsten von Bergmann  and André Kubetzka  and Roland Wiesendanger },
	title = {Writing and Deleting Single Magnetic Skyrmions},
	journal = {Science},
	volume = {341},
	number = {6146},
	pages = {636-639},
	year = {2013},
	doi = {10.1126/science.1240573},
}

@article{jiang2015blowing,
	author = {Wanjun Jiang  and Pramey Upadhyaya  and Wei Zhang  and Guoqiang Yu  and M. Benjamin Jungfleisch  and Frank Y. Fradin  and John E. Pearson  and Yaroslav Tserkovnyak  and Kang L. Wang  and Olle Heinonen  and Suzanne G. E. te Velthuis  and Axel Hoffmann },
	title = {Blowing magnetic skyrmion bubbles},
	journal = {Science},
	volume = {349},
	number = {6245},
	pages = {283-286},
	year = {2015},
	doi = {10.1126/science.aaa1442},
}

@article{fert2017magnetic,
	title={Magnetic skyrmions: advances in physics and potential applications},
	author={Fert, Albert and Reyren, Nicolas and Cros, Vincent},
	journal={Nat. Rev. Mater.},
	volume={2},
	number={7},
	pages={1--15},
	year={2017},
	publisher={Nature Publishing Group},
	doi = {10.1038/natrevmats.2017.31},
}

@article{bogdanov2020physical,
	title={Physical foundations and basic properties of magnetic skyrmions},
	author={Bogdanov, Alexei N and Panagopoulos, Christos},
	journal={Nat. Rev. Phys.},
	volume={2},
	number={9},
	pages={492--498},
	year={2020},
	publisher={Nature Publishing Group UK London},
	doi = {10.1038/s42254-020-0203-7},
}

@article{reichhardt2022statics,
	title = {Statics and dynamics of skyrmions interacting with disorder and nanostructures},
	author = {Reichhardt, C. and Reichhardt, C. J. O. and Milo\ifmmode \check{s}\else \v{s}\fi{}evi\ifmmode \acute{c}\else \'{c}\fi{}, M. V.},
	journal = {Rev. Mod. Phys.},
	volume = {94},
	issue = {3},
	pages = {035005},
	numpages = {61},
	year = {2022},
	month = {Sep},
	publisher = {American Physical Society},
	doi = {10.1103/RevModPhys.94.035005},
}

@article{science.aau0227,
	author = {S. Tsesses  and E. Ostrovsky  and K. Cohen  and B. Gjonaj  and N. H. Lindner  and G. Bartal },
	title = {Optical skyrmion lattice in evanescent electromagnetic fields},
	journal = {Science},
	volume = {361},
	number = {6406},
	pages = {993-996},
	year = {2018},
	doi = {10.1126/science.aau0227},
}

@article{du2019deep,
	title={Deep-subwavelength features of photonic skyrmions in a confined electromagnetic field with orbital angular momentum},
	author={Du, Luping and Yang, Aiping and Zayats, Anatoly V and Yuan, Xiaocong},
	journal={Nat. Phys.},
	volume={15},
	number={7},
	pages={650--654},
	year={2019},
	publisher={Nature Publishing Group UK London},
	doi = {10.1038/s41567-019-0487-7},
}

@article{science.aba6415,
	author = {Timothy J. Davis  and David Janoschka  and Pascal Dreher  and Bettina Frank  and Frank-J. Meyer zu Heringdorf  and Harald Giessen},
	title = {Ultrafast vector imaging of plasmonic skyrmion dynamics with deep subwavelength resolution},
	journal = {Science},
	volume = {368},
	number = {6489},
	pages = {eaba6415},
	year = {2020},
	doi = {10.1126/science.aba6415},
}

@article{PhysRevLett.127.144502,
	title = {Observation of Acoustic Skyrmions},
	author = {Ge, Hao and Xu, Xiang-Yuan and Liu, Le and Xu, Rui and Lin, Zhi-Kang and Yu, Si-Yuan and Bao, Ming and Jiang, Jian-Hua and Lu, Ming-Hui and Chen, Yan-Feng},
	journal = {Phys. Rev. Lett.},
	volume = {127},
	issue = {14},
	pages = {144502},
	numpages = {6},
	year = {2021},
	month = {Sep},
	publisher = {American Physical Society},
	doi = {10.1103/PhysRevLett.127.144502},
}

@article{sciadv.adf3652,
	author = {Liyun Cao  and Sheng Wan  and Yi Zeng  and Yifan Zhu  and Badreddine Assouar },
	title = {Observation of phononic skyrmions based on hybrid spin of elastic waves},
	journal = {Sci. Adv.},
	volume = {9},
	number = {7},
	pages = {eadf3652},
	year = {2023},
	doi = {10.1126/sciadv.adf3652},
}

@article{wang2025topological,
	title={Topological water-wave structures manipulating particles},
	author={Wang, Bo and Che, Zhiyuan and Cheng, Cheng and Tong, Caili and Shi, Lei and Shen, Yijie and Bliokh, Konstantin Y and Zi, Jian},
	journal={Nature},
	volume = {638},
	pages={394--400},
	year={2025},
	publisher={Nature Publishing Group UK London},
	doi = {10.1038/s41586-024-08384-y},
}

@article{PhysRevResearch.3.023055,
	title = {Microcavity-based generation of full Poincar\'e beams with arbitrary skyrmion numbers},
	author = {Lin, Wenbo and Ota, Yasutomo and Arakawa, Yasuhiko and Iwamoto, Satoshi},
	journal = {Phys. Rev. Res.},
	volume = {3},
	issue = {2},
	pages = {023055},
	numpages = {12},
	year = {2021},
	month = {Apr},
	publisher = {American Physical Society},
	doi = {10.1103/PhysRevResearch.3.023055},
}

@article{Kerridge-Johns:24,
	author = {William R. Kerridge-Johns and A. Srinivasa Rao and Takashige Omatsu},
	journal = {Optica},
	number = {6},
	pages = {769--775},
	publisher = {Optica Publishing Group},
	title = {Optical skyrmion laser using a wedged output coupler},
	volume = {11},
	month = {Jun},
	year = {2024},
	doi = {10.1364/OPTICA.521901},
}

@article{wang2025perturbation,
	title={Perturbation-resilient integer arithmetic using optical skyrmions},
	author={Wang, An Aloysius and Ma, Yifei and Zhang, Yunqi and Zhao, Zimo and Cai, Yuxi and Qiu, Xuke and Dong, Bowei and He, Chao},
	journal={Nat. Photon.},
	volume = {19},
	pages={1367--1375},
	year={2025},
	publisher={Nature Publishing Group UK London},
	doi = {10.1038/s41566-025-01779-x},
}

@article{chen2025programmable,
	title={Programmable skyrmions for robust communication and intelligent sensing},
	author={Chen, Long and Li, Xin Yu and Shen, Yijie and Gu, Ze and Su, Jian Lin and Xiao, Qiang and Huang, Si Qi and Qin, Shi Long and Ma, Qian and You, Jian Wei and others},
	journal={arXiv:2507.05944},
	year={2025},
	url = {https://doi.org/10.48550/arXiv.2507.05944},
}

@article{jiang2017direct,
	title={Direct observation of the skyrmion {Hall} effect},
	author={Jiang, Wanjun and Zhang, Xichao and Yu, Guoqiang and Zhang, Wei and Wang, Xiao and Benjamin Jungfleisch, M and Pearson, John E and Cheng, Xuemei and Heinonen, Olle and Wang, Kang L and others},
	journal={Nat. Phys.},
	volume={13},
	number={2},
	pages={162--169},
	year={2017},
	publisher={Nature Publishing Group UK London},
	doi = {10.1038/nphys3883},
}

@article{litzius2017skyrmion,
	title={Skyrmion {Hall} effect revealed by direct time-resolved X-ray microscopy},
	author={Litzius, Kai and Lemesh, Ivan and Kr{\"u}ger, Benjamin and Bassirian, Pedram and Caretta, Lucas and Richter, Kornel and B{\"u}ttner, Felix and Sato, Koji and Tretiakov, Oleg A and F{\"o}rster, Johannes and others},
	journal={Nat. Phys.},
	volume={13},
	number={2},
	pages={170--175},
	year={2017},
	publisher={Nature Publishing Group UK London},
	doi = {10.1038/nphys4000}
}

@article{gobel2021beyond,
	title = {Beyond skyrmions: Review and perspectives of alternative magnetic quasiparticles},
	journal = {Phys. Rep.},
	volume = {895},
	pages = {1-28},
	year = {2021},
	doi = {10.1016/j.physrep.2020.10.001},
	author = {Börge Göbel and Ingrid Mertig and Oleg A. Tretiakov},
}

@article{PhysRevA.102.053513,
	title = {Paraxial skyrmionic beams},
	author = {Gao, Sijia and Speirits, Fiona C. and Castellucci, Francesco and Franke-Arnold, Sonja and Barnett, Stephen M. and G\"otte, J\"org B.},
	journal = {Phys. Rev. A},
	volume = {102},
	issue = {5},
	pages = {053513},
	numpages = {6},
	year = {2020},
	month = {Nov},
	publisher = {American Physical Society},
	doi = {10.1103/PhysRevA.102.053513},
}

@article{shen2021supertoroidal,
	title={Supertoroidal light pulses as electromagnetic skyrmions propagating in free space},
	author={Shen, Yijie and Hou, Yaonan and Papasimakis, Nikitas and Zheludev, Nikolay I},
	journal={Nat. Commun.},
	volume={12},
	number={1},
	pages={5891},
	year={2021},
	publisher={Nature Publishing Group UK London},
	doi = {10.1038/s41467-021-26037-w},
}

@article{teng2025construction,
	title={Construction of optical spatiotemporal skyrmions},
	author={Teng, Houan and Liu, Xin and Zhang, Nianjia and Fan, Haihao and Chen, Guoliang and Cao, Qian and Zhong, Jinzhan and Lei, Xinrui and Zhan, Qiwen},
	journal={Light. Sci. Appl.},
	volume={14},
	number={1},
	pages={324},
	year={2025},
	publisher={Nature Publishing Group UK London},
	doi = {10.1038/s41377-025-02028-0},
}

@article{fan2025topological,
	title={Topological state and number transitions of optical skyrmions upon free-space beam propagation},
	author={Fan, Xinhao and Wu, Xuanguang and Ren, Kang and Zhou, Liang and Guo, Xuyue and Wei, Bingyan and Zhang, Yi and Liu, Sheng and Li, Peng and Zhao, Jianlin},
	journal={Commun. Phys.},
	year={2025},
	volume={8},
	pages={500},
	publisher={Nature Publishing Group UK London},
	doi = {10.1038/s42005-025-02427-0},
}

@article{Zhen26,
	author = {Weiming Zhen and Zhiming Qing and Wenxiang Yan and Zhi-Cheng Ren and Xi-Lin Wang and Hui-Tian Wang and Jianping Ding and Yijie Shen},
	journal = {Optica},
	number = {2},
	pages = {188--194},
	publisher = {Optica Publishing Group},
	title = {Reconfiguring optical skyrmion topology in free space},
	volume = {13},
	month = {Feb},
	year = {2026},
	doi = {10.1364/OPTICA.579220},
}

@article{sun2024acoustic,
	title={Acoustic skyrmionic mode coupling and transferring in a chain of subwavelength metastructures},
	author={Sun, Wen-Jun and Zhou, Nong and Chen, Wan-Na and Sheng, Zong-Qiang and Wu, Hong-Wei},
	journal={Adv. Sci.},
	volume={11},
	number={34},
	pages={2401370},
	year={2024},
	publisher={Wiley Online Library},
	doi = {10.1002/advs.202401370},
}

@article{PhysRevB.80.125327,
	title = {Robustness of the {spin-Chern} number},
	author = {Prodan, Emil},
	journal = {Phys. Rev. B},
	volume = {80},
	issue = {12},
	pages = {125327},
	numpages = {7},
	year = {2009},
	month = {Sep},
	publisher = {American Physical Society},
	doi = {10.1103/PhysRevB.80.125327},
}

@article{PhysRevLett.107.066602,
	title = {Time-Reversal-Symmetry-Broken Quantum Spin {Hall} Effect},
	author = {Yang, Yunyou and Xu, Zhong and Sheng, L. and Wang, Baigeng and Xing, D. Y. and Sheng, D. N.},
	journal = {Phys. Rev. Lett.},
	volume = {107},
	issue = {6},
	pages = {066602},
	numpages = {5},
	year = {2011},
	month = {Aug},
	publisher = {American Physical Society},
	doi = {10.1103/PhysRevLett.107.066602},
}

@article{deng2020acoustic,
	title={Acoustic {spin-Chern} insulator induced by synthetic spin--orbit coupling with spin conservation breaking},
	author={Deng, Weiyin and Huang, Xueqin and Lu, Jiuyang and Peri, Valerio and Li, Feng and Huber, Sebastian D and Liu, Zhengyou},
	journal={Nat. Commun.},
	volume={11},
	number={1},
	pages={3227},
	year={2020},
	publisher={Nature Publishing Group UK London},
	doi = {10.1038/s41467-020-17039-1},
}

@article{advs.202409377,
	author = {Liu, Lei and Sun, Xiao-Chen and Tian, Yuan and Zhang, Xiujuan and Lu, Ming-Hui and Chen, Yan-Feng},
	title = {Cyclic Evolution of Synergized Spin and Orbital Angular Momenta},
	journal = {Adv. Sci.},
	volume = {12},
	number = {3},
	pages = {2409377},
	doi = {10.1002/advs.202409377},
	year = {2025}
}

@article{liu2025acoustic,
	title={Acoustic spin skyrmion molecule lattices enabling stable transport and flexible manipulation},
	author={Liu, Lei and Zhang, Xiujuan and Lu, Ming-Hui and Chen, Yan-Feng},
	journal={Nat. Commun.},
	volume={16},
	number={1},
	pages={10607},
	year={2025},
	publisher={Nature Publishing Group UK London},
	doi = {10.1038/s41467-025-65611-4},
}

@article{PhysRevLett.128.073602,
	title = {Perfect Chirality with Imperfect Polarization},
	author = {Lang, Ben and McCutcheon, Dara P. S. and Harbord, Edmund and Young, Andrew B. and Oulton, Ruth},
	journal = {Phys. Rev. Lett.},
	volume = {128},
	issue = {7},
	pages = {073602},
	numpages = {6},
	year = {2022},
	month = {Feb},
	publisher = {American Physical Society},
	doi = {10.1103/PhysRevLett.128.073602},
}

@misc{supp,
	note = "See Supplemental Material at URL for further details."
}

@dataset{liu_2026_18328890,
	author       = {Liu, Lei},
	title        = {Experimental data for Acoustic quantum skyrmion-valley {Hall} effect},
	month        = {jan},
	year         = {2026},
	publisher    = {Zenodo},
	doi          = {10.5281/zenodo.18328890}
}
\end{document}